\documentclass{aa}
\usepackage{graphicx}
\usepackage{txfonts}
\usepackage{color}

\begin{document}
\definecolor{red}{rgb}{1,0,0}
\definecolor{blue}{rgb}{0,0,1}
%
\title{The molecular and dusty composition of Betelgeuse's inner circumstellar environment
\thanks{Based on observations collected at ESO with the VLTI/MIDI instrument.}}

\author{G. Perrin \inst{1}
\and T. Verhoelst  \inst{2}  \fnmsep \thanks{Postdoctoral Fellow of
the Fund for Scientific Research, Flanders}
\and S.T. Ridgway \inst{3} 
\and J. Cami \inst{4}
\and Q.N. Nguyen \inst{1}
\and O. Chesneau \inst{5}
\and B. Lopez \inst{5}
\and Ch. Leinert \inst{6}
\and A. Richichi \inst{7}}

   \offprints{G. Perrin, \email{guy.perrin@obspm.fr}}

\institute{Observatoire de Paris, LESIA, UMR 8109, F-92190 Meudon, 
France \and  Instituut voor Sterrenkunde, K.U. Leuven, Celestijnenlaan 200D, B-3001
Leuven, Belgium
  \and National Optical Astronomy Observatories, Tucson, AZ, USA
  \and Physics \& Astronomy Dept., University of Western Ontario, London ON  N6A 3K7, Canada
  \and Observatoire de la C\^ote dÕAzur, G\'emini, UMR 6302, BP 4229, F-06304 Nice, France
  \and Max Planck Institute for Astronomy, K\"onigstuhl 17, 69117 Heidelberg, Germany
  \and European Southern Observatory, Karl-Schwarzschildstr. 2, 85748 Garching bei M\"unchen, Germany}

   \date{Received ; accepted }

  
    \abstract
   {The study of the atmosphere of red supergiant stars in general and of Betelgeuse ($\alpha$ Orionis) in particular is of prime importance to understand dust formation and how mass is lost to the interstellar medium in evolved massive stars.}
   {A molecular shell, the MOLsphere (Tsuji, 2000a), in the atmosphere of Betelgeuse has been proposed to account for the near- and mid-infrared spectroscopic observations of Betelgeuse. The goal is to further test this hypothesis and to identify some of the molecules in this MOLsphere.}
   {We report on measurements taken with the mid-infrared two-telescope beam combiner of the VLTI, MIDI, operated between 7.5 and 13.5~$\mu$m. The data are compared to a simple geometric model of a photosphere surrounded by a warm absorbing and emitting shell. Physical characteristics of the shell are derived: size, temperature and optical depth. The chemical constituents are determined with an analysis consistent with available infrared spectra and interferometric data. }
   {The MIDI data are well modeled with a geometrically thin shell whose radius varies from 1.31 to 1.43\,$R_\star$ across the N band with a typical temperature of 1550\,K. We are able to account for the measured optical depth of the shell in the N band, the ISO-SWS spectrum and K and L band interferometric data with a shell whose inner and outer radii are given by the above range and with the following species and densities: H$_2$O ($7.1 \pm 4.7 \times 10^{19}$\,cm$^{-2}$), 
   	SiO ($4.0 \pm 1.1 \times 10^{20}$\,cm$^{-2}$), Al$_2$O$_3$ ($2.4 \pm 0.5 \times 10^{15}$\,cm$^{-2}$).}
   {These results confirm the MOLsphere model. We bring evidence for more constituents and for the presence of species participating in the formation of dust grains in the atmosphere of the star, i.e. well below the distance at which the dust shell is detected. We believe these results bring key elements to the understanding of mass loss in Betelgeuse and red supergiants in general and bring support to the dust-driven scenario. }

   \maketitle

\section{Introduction}
The atmospheres of late-type supergiant stars are not {\it a priori} supported by large amplitude pulsations of the star as is the case in Mira stars but are still much more extended than expected in hydrostatic equilibrium. In the case of Betelgeuse, this is directly illustrated by apparent sizes ranging from 42 up to 65~mas, the size being minimum in the near-infrared and larger in the visible and ultraviolet. The large extent of the atmosphere is not understood and makes it complex to study.

Strong water-vapor bands have been detected in $\mu$~Cep and Betelgeuse by \cite{danielson1965} but no explanation of their origin was available at that time. Ê\cite{jennings1998} detected pure rotational lines of H$_2$O at 12.3 $\mu$m in Betelgeuese and Antares  and suggested that the water lines may be formed in their atmosphere.
Tsuji (2000a, 2000b) reinterpreted previously published data (Wing \& Spinrad 1970) and proposed a $1500\pm500$~K water vapor and CO shell around Betelgeuse and $\mu$~Cep, the MOLsphere.
 
Multi-telescope optical interferometry offers new and important constraints on the spatial location of these regions inferred from spectroscopy. \cite{perrin2004a}  have shown that the near-infrared structure of Betelgeuse can be described by two main components: a photosphere of temperature $3641\pm53$~K and a gaseous shell at a temperature of $2055\pm25$~K and located $0.33~R_{\star}$ above the photosphere. The model proved to consistently explain the FLUOR/IOTA K-band measurements and, after removing the contribution of dust, those of the  ISI interferometer at $11.15$~$\mu$m of \cite{weiner2000}. The same conclusion was reached on $\mu$~Cep based on narrow band measurements in the K band to isolate molecular bands from the continuum. The shell has been found to have a $1.32~R_{\star}$ radius (Perrin et al. 2005). 
This analysis was also successfully applied to Mira stars (Mennesson et al. 2002, Perrin et al. 2004b) and led to a consistent understanding of seemingly contradictory measurements from the visible to the near infrared. \\
\indent Other recent modeling studies of supergiants (Ohnaka 2004a) and Mira stars (Ohnaka 2004b, Weiner 2004) have shown that the near- and mid-infrared H$_{2}$O spectra and the apparent near- and mid-infrared angular sizes of both stellar types can be consistently understood in terms of a star with an envelope of warm H$_{2}$O vapor.  These results altogether confirm and vindicate Tsuji's MOLsphere. In a more recent work,  \cite{tsuji2006} shows the consistency of  interferometry and spectroscopy-based conclusions. \cite{verhoelst2006} have made a combined analysis of the near and mid-infrared undispersed interferometric and spectroscopic data and showed in addition to H$_2$O, OH and CO the likelihood of amorphous alumina in the MOLsphere, a dust species with a high sublimation temperature. \cite{ryde2006} proposes an explanation for mid-infrared TEXES spectroscopic data which does not require a MOLsphere. Absorption features of water vapor and OH can be reproduced with a cooler outer photospheric structure of temperature 3250~K. However, this is not compatible with near-infrared spectra according to \cite{tsuji2006}.

\begin{table}[htbp]
      \caption[]{Reference sources. Diameters are based on photometric estimates by \cite{dyck1996} with errors revised by \cite{perrin1998}. Although not directly used for calibration, $\beta$~Lep and $\eta$~Eri have been used for data selection.}
         \label{tab:cal}
         \begin{tabular}{lccc}
            \hline
            \hline
            \noalign{\smallskip}
            HD number & Source  &  Spectral type & Uniform disk  \\
	                 & name &                & diameter (mas)\\
            \noalign{\smallskip}
            \hline
	HD 39400	& 56  Ori		& K1.5 IIb	& $2.25\pm0.12$ \\
	HD 49161	& 17 Mon		& K4 III	& $2.65\pm0.13$ \\
	HD 36079	& $\beta$ Lep	& G5 II	& $3.04\pm0.15$ \\
	HD 18322	& $\eta$ Eri	& K1 III	& $2.57\pm0.13$ \\
          \noalign{\smallskip}
            \noalign{\smallskip}
            \hline
         \end{tabular}
   \end{table}

Although excellent candidates are known as potential contributors to the opacity of the MOLsphere, other sources may be considered such as electron-hydrogen collisions as proposed by \cite{tatebe2006} for the shell around Mira stars. The combination of spectroscopy and interferometry will help in this research. In the present paper, spectrally dispersed fringes in the N band obtained with MIDI at VLTI allow to evaluate the compact spherical shell model for a new wavelength range. We first present the observations in Sect.~\ref{data} and discuss the data reduction. Visibility data are fitted with a simple uniform disk + extended resolved source model in Sect.~\ref{sec:diameter} to yield normalized visibilities free of any extended component. The compact source visibilities are then compared to a star + molecular spherical shell model in Sect.~\ref{model}. The shell optical depths are compared to a spectroscopic model in Sect.~\ref{spectro} to derive column densities and identify molecules from their signatures. Lastly, the results and their consequences are discussed in Sect.~\ref{discussion}.


\section{Observations and data reduction} 
\label{data}  
          
\begin{table*}[htbp]
\caption[]{Log of November 8, 2003 observations.}
\label{tab:V^2}
\begin{center}
\begin{tabular}{cccclclc}
\hline
\hline
\noalign{\smallskip}
UT time	& Projected baseline	& Azimuth$^{\mathrm{a}}$ & opd offset	& Calibrator 1	& UT time	& Calibrator 2	& UT time \\
		&  (m)				& ($\degr$) &		($\mu$m) & & 	&	& \\
\noalign{\smallskip}
\hline
\noalign{\smallskip}

06:11:07.00	& 37.320050	& 38.794453	&	250  & 56 Ori 	& 05:29:14.43	& 56 Ori	& 06:40:55.00 \\

06:12:07.28	& 37.396414	& 38.907380	&       185 & 56 Ori	& 05:29:14.43	& 56 Ori	& 06:40:55.00 \\


06:16:01.20 	& 37.691753	& 39.335562	&       185  & 56 Ori	& 05:29:14.43	& 56 Ori	& 06:40:55.00 \\
	    
07:10:06.99	& 41.509800	& 43.801921	&        180 & 56 Ori	& 06:42:11.45	& 17 Mon	& 07:43:28.00 \\

07:10:59.20 	& 41.565253	& 43.853488	&        180  & 56 Ori	& 06:42:11.45	& 17 Mon	& 07:43:28.00 \\

08:09:18.00	& 44.673489	& 46.091170	&        170 & 17 Mon		& 07:44:44.45	&		& \\

08:10:10.20 	& 44.709776	& 46.107820	&         170 & 17 Mon		& 07:44:44.45	&		& \\

\noalign{\smallskip}
\hline
\end{tabular}
\end{center}
\begin{list}{}{}
\item[$^{\mathrm{a}}\,\,$counted positive from East towards North]
\end{list}
\end{table*}

Observations took place at Cerro Paranal in Chile and made use of the mid-infrared beamcombiner MIDI (Leinert et al. 2003) of the VLTI on the nights of November 8 and 10, 2003. These observations were part of a Science Demonstration Time run and made use of the UT2-UT3 baseline whose 46.63~m length allows to sample the first zero region of the visibility function of Betelgeuse. 

The MIDI data comprise different sets, two of which are important for data reduction. The chopping set provides the average flux of the source from which the strong background contribution has been removed. The fringe set is a collection of about a hundred 50~$\mu$m long optical path difference scans. At the time these observations took place, the MIDI beams were not spatially filtered. A 0.6~arcsec slit was used corresponding to a $2.5\lambda/D$ scale at the central wavelength of 10~$\mu$m. The beams - the two complementary interferometric outputs of the beamcombiner - were dispersed by a prism at a spectral resolution of 30 in the direction perpendicular to the slit. No active system was used to correct wavefronts except for the tip-tilt correction system to maintain the pointing quality. Given that pupils are larger than the coherence patch of atmospheric turbulent phase (about 6~m for median seeing at Paranal at 10~$\mu$m) and that the beams were not filtered, coherence losses due to turbulence fluctuate from one scan to another and may prove difficult to calibrate. This is the major source of error in measuring visibilities as is discussed below.
 
An interferogram is the sum of three signals: the background, the flux of the source and the fringe modulation. Because the beams are not filtered and scintillation can be neglected at 10~$\mu$m with an 8.2~m telescope, the flux of the source is constant throughout a batch and can be accurately measured with the chopping sequence. The quick and large fluctuations of the background prevent from simultaneously measuring the source flux on interferograms with the MIDI set-up. As a consequence, the fringe modulation has to be normalized with the estimate of the flux of the source derived from the chopping sequences. To do so, the two complementary interferometric outputs are subtracted to eliminate the background and the instantaneous source flux leaving a zero mean signal. The power spectral density of normalized interferograms is integrated to yield the squared coherence factor of each  interferogram in a batch. The power spectral density of the sum of noises (background photon noise, source photon noise and detector read-out noise) has been subtracted before the integration to remove the bias on visibility modulus due to additive noises. The average of all squared coherence factors in a batch yields an estimate of the squared coherence factor for the batch and their dispersion yields an estimate of the attached uncertainty. A first selection of data occurs here. Data are rejected when squared coherence factor histograms strongly depart from a gaussian distribution as this is the signature of quick evolution of turbulence statistics during a batch thus strongly biasing the average value estimate. 

This procedure is applied to both Betelgeuse and calibrator star data. Calibrator star squared coherence factors are divided by expected squared visibilities estimated from their uniform disk diameter value to compute the squared transfer function. When observations of Betelgeuse are bracketted by two calibrator observations, transfer functions are interpolated at the time Betelgeuse was observed as explained in \cite{perrin2003} and error bars are computed accordingly. The final squared visibility estimate of the source is the ratio of the squared coherence factor and of the transfer function and error bars are derived from the errors on these two quantities. This procedure is applied to each individual spectral channel.  It is simpler than what is usually done in the near-infrared where turbulence fluctuations are filtered with single-mode fibers. Non stationary errors due to turbulence are impossible to accurately calibrate here and there is no need for a sophisticated propagation of noises and uncertainties. 

The calibration procedure is applied after data have been checked for quality. Transfer functions are plotted as a function of wavelength for each calibrator. Ideally they should all overlap to within noise. Transfer functions are rejected when they clearly depart from the average trend. This can be easily detected in as much as a sudden increase of turbulence or a bad image overlap will cause the transfer function to drop and to drop all the more as the wavelength is shorter. Once the selection of transfer functions is performed, Betelgeuse data are calibrated and the same procedure applies to visibility curves as a function of wavelength. The application of this procedure has led to reject all November 10, 2003 data and part of November 8, 2003.

All the scans of the first night on Betelgeuse were recorded off the zero optical path difference. Fortunately, in dispersed mode, the spectral resolution is high enough that the offset is within the coherence length of the radiation. Most offsets were in the range 170-180~$\mu$m with one batch with an offset of 250~$\mu$m (see Table~\ref{tab:V^2}). In the worst case, the offset caused a contrast reduction of 86\% at 8~$\mu$m and 14\% at 13~$\mu$m. But for the other batches, the contrast reduction was 60\% at 8~$\mu$m and 7\% at 13~$\mu$m. In order to calibrate these losses, we simulated the interferograms recorded by each pixel of the MIDI detector taking into account the dispersion law of the prism and the effects of diffraction in the MIDI spectrograph. Losses only depend on the characteristics of the instrument and can therefore be reliably computed. Laws of contrast loss as a function of wavelength were derived for each optical path difference offset and used to calibrate all acquisitions. All the calibrator scans were recorded at zero optical path difference and no correction was necessary for calibrators. 

\begin{table*}[htbp]
\caption[]{November 8, 2003 averaged observations.}
\label{tab:bin}
\begin{center}
\begin{tabular}{ccclclc}
\hline
\hline
\noalign{\smallskip}
Mean UT time	& Mean projected baseline	& Mean azimuth$^{\mathrm{a}}$	& Calibrator 1	& UT time	& Calibrator 2	& UT time \\
		&  (m)				& ($\degr$) &		&	&	& \\
\noalign{\smallskip}
\hline
\noalign{\smallskip}

06:13:36.12	& 37.508548	& 39.906969	&	56 Ori 	& 05:29:14.43	& 56 Ori	& 06:40:55.00 \\

07:10:33.09	& 41.537526	& 43.827705	&  56 Ori	& 06:42:11.45	& 17 Mon	& 07:43:28.00 \\

08:09:44.10	& 44.691632	& 46.099496	& 17 Mon		& 07:44:44.45	&		& \\

\noalign{\smallskip}
\hline
\end{tabular}
\end{center}
\begin{list}{}{}
\item[$^{\mathrm{a}}\,\,$counted positive from East towards North]
\end{list}
\end{table*}

The entire data reduction has made use of the software developed by Paris Observatory for MIDI\footnote{Software is available through the JMMC website: http://mariotti.ujf-grenoble.fr/}. The list of calibrators and the log of calibrated data are given in Table~\ref{tab:cal} and Table~\ref{tab:V^2}. Three different sets of projected baselines have been observed. We have averaged data  to get three bins corresponding to the sets. Averaging has been performed using the statistical errors provided by the data reduction software. Each average is a $\chi^2$ estimator:
\begin{equation}
\chi^2=\frac{1}{N-1}\sum_{i=0}^{N-1}\frac{(V^2_{\mathrm{bin}}-V^2_i)^2}{\sigma_i^2}
\end{equation}
The statistical variance of the average visibility $V^2_{\mathrm{bin}}$ is:
\begin{equation}
\sigma^2=\left[{\sum_{i=0}^{N-1}1/\sigma_i^2}\right]^{-1}
\end{equation}
The $\chi^2$ can be used as an indicator of the correctness of error bars and is useful to improve them. If the $\chi^2$ is smaller than 1 then statistical error bars are declared correct and $\sigma^2_{\mathrm{bin}}=\sigma^2$. Otherwise, the error bar on the averaged visibility is corrected applying:
\begin{equation}
\sigma^2_{\mathrm{bin}}=\chi^2*\sigma^2
\end{equation}
which is equivalent to forcing the $\chi^2$ at minimum to 1.This procedure is meant to account for non stationary fluctuations of turbulent errors which are not detected by the data reduction software. This does not perfectly remove biases as these are not modeled but at least error bars are increased to better model fluctuations. Typical error bars are a fraction of a percent. The validity of this approach is assessed in the next section. Given the normalization we had to apply to the data, it is difficult to estimate the validity of our error bars as absolute errors. However, the derived error bars are internally consistent. The log of averaged observations is given in Table~\ref{tab:bin}.


\section{Diameter fit}
\label{sec:diameter}
\begin{figure*}[]
\hbox{
   \includegraphics[bb=60 58 565 775 , angle=90, width=8.95cm]{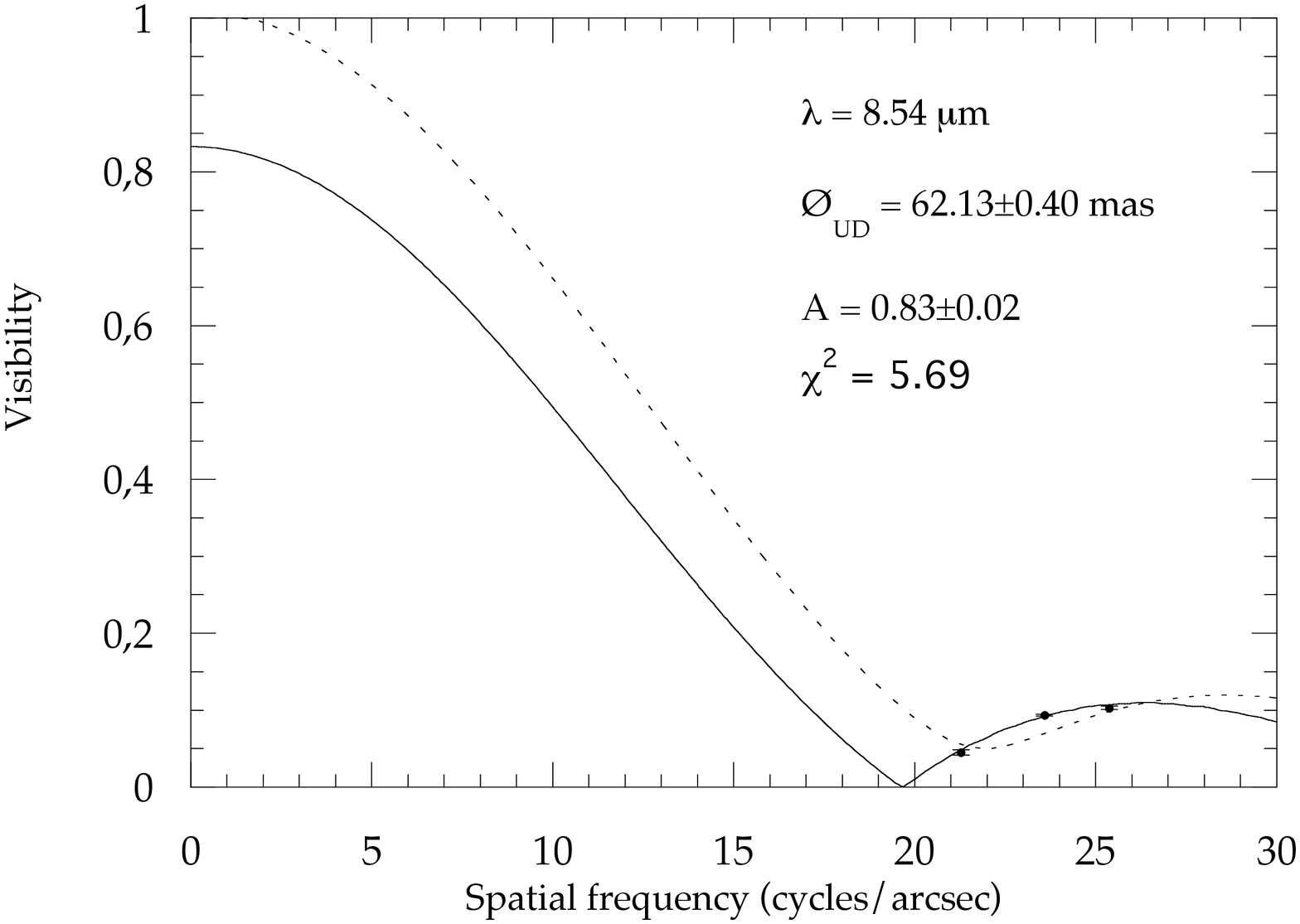}
   \includegraphics[bb=60 58 565 775 , angle=90, width=8.95cm]{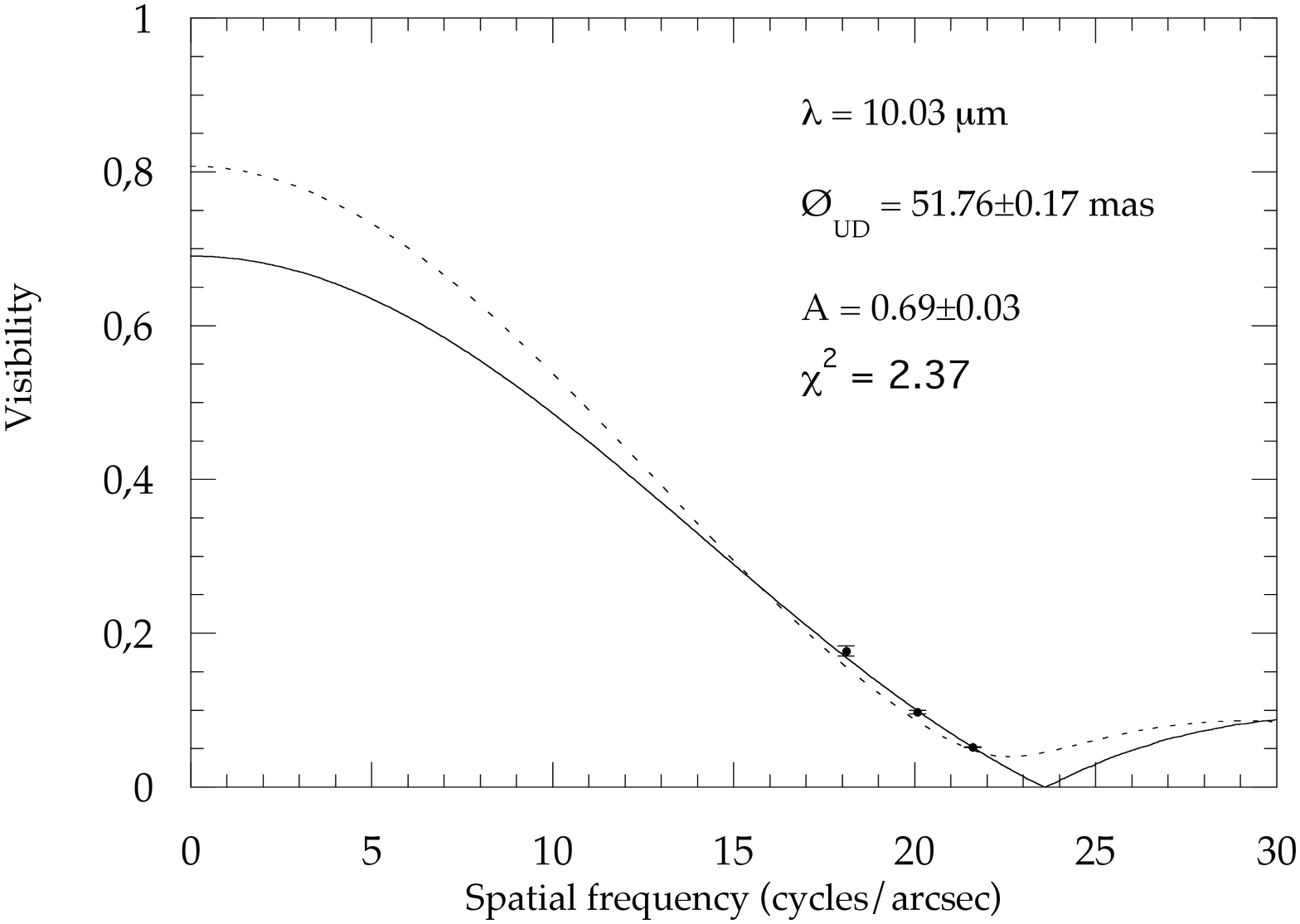}
} 
\vspace{1mm}
\hbox{
   \includegraphics[bb=60 58 565 775 , angle=90, width=8.95cm]{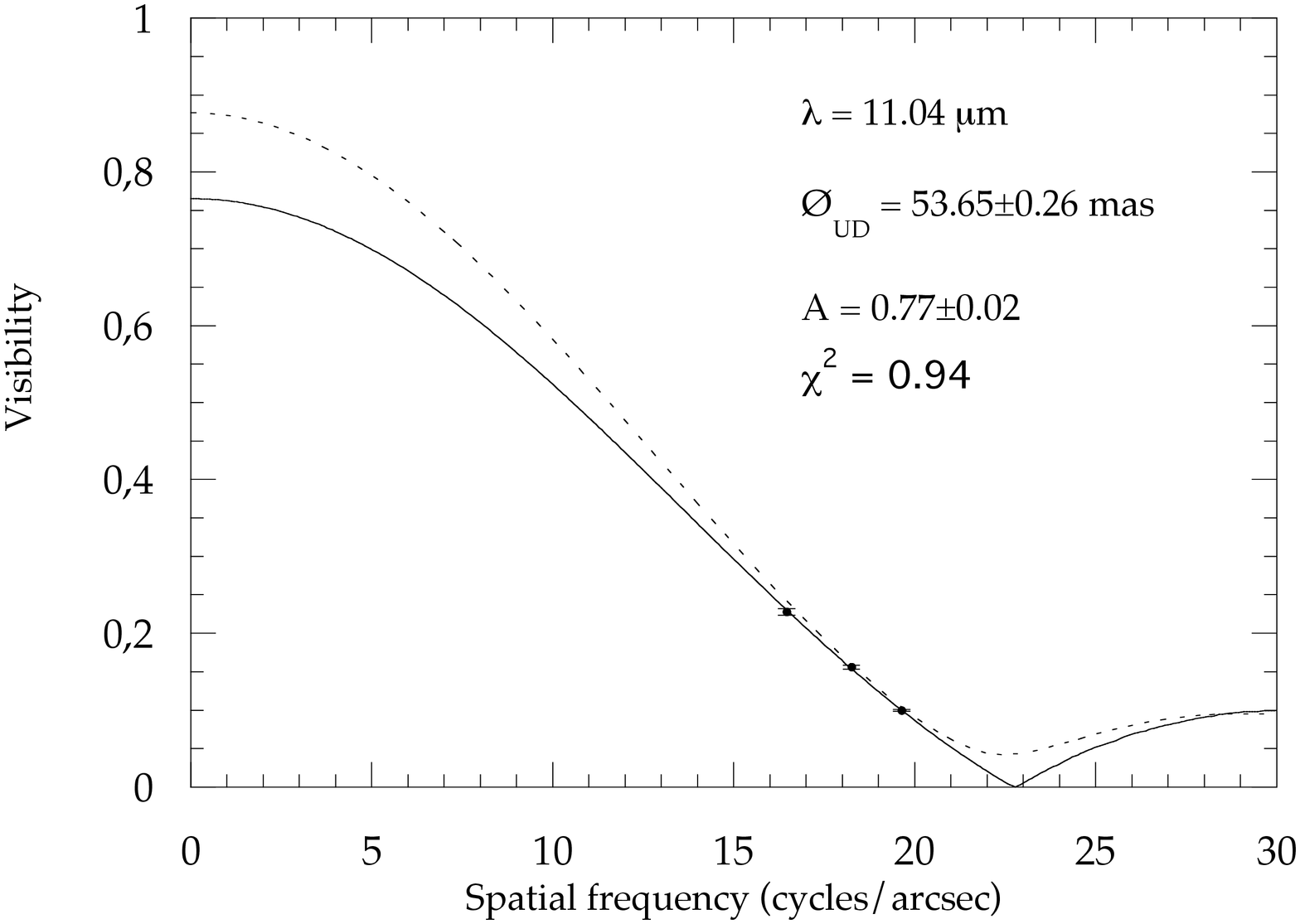}
   \includegraphics[bb=60 58 565 775 , angle=90, width=8.95cm]{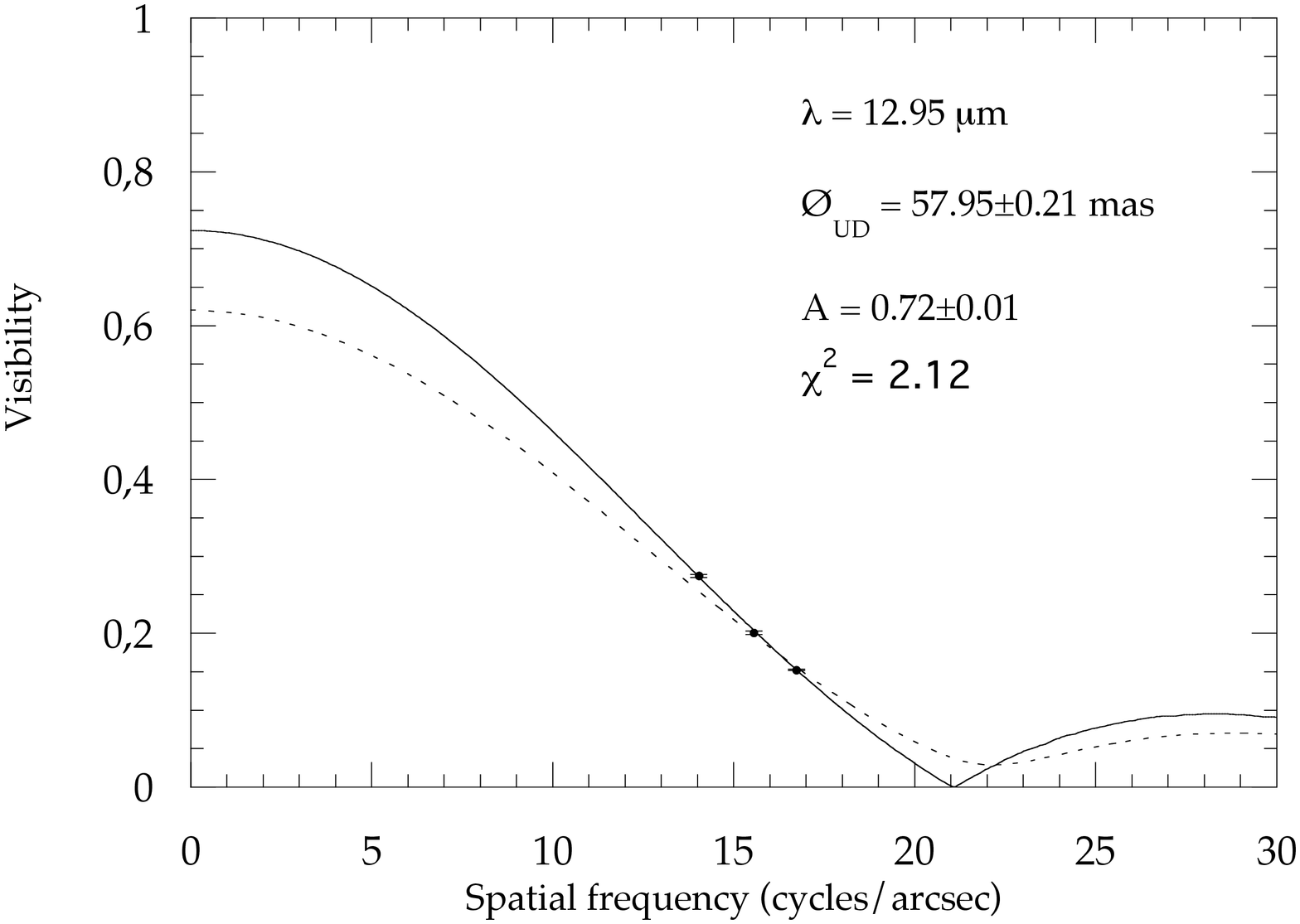}
}   
\vspace{2mm}
     \caption{Examples of MIDI visibility fits by the scaled uniform disk visibility function (solid line). Fits with the spectroscopic model of Sect.~\ref{direct} with a 1520~K layer are also presented (dashed line).}
         \label{fig:fits}
   \end{figure*}
Betelgeuse is known to be surrounded by a dust envelope whose characteristic size of 1\arcsec  (Danchi et al. 1994) is far larger than the interfering field of MIDI and decreases the spatial coherence of the beams. To first approximation, the source can be modeled with an extended and a compact component. The extended component leads to a decrease of the compact component visibility by a factor equal to its fractional flux. The study reported here only addresses the compact object. We therefore need to normalize visibilities to clean the compact source from the disturbing extended component. 

Visibilities in all spectral channels have been fitted by the scaled uniform disk function:
\begin{equation}
V^2(\O_{UD}(\lambda),A,\lambda;B_{i})=
A^2{\left|\frac{2J_{1}\left(\pi\O_{UD}\frac{B_{i}}{\lambda}\right)}{\pi\O_{UD}\frac{B_{i}}{\lambda}}\right|}^{2}
\label{eq:vis}
\end{equation}
where A is the fractional flux of the compact component, $\O_{UD}(\lambda)$ its diameter and $B_i$ the projected baseline. We assume here that the visibility of the extended component is equal to 0. Visibilities have been taken close to the first null of the compact source visibility function. The effect of finite bandwidth has to be taken into account as the estimated squared visibilities do not go through 0. For comparison with observations, the above model has therefore been integrated in each spectral channel as described in \cite{perrin2004a}.

\begin{figure*}[t]
\hbox{
   \includegraphics[bb=60 58 565 775 , angle=-90, width=8.95cm]{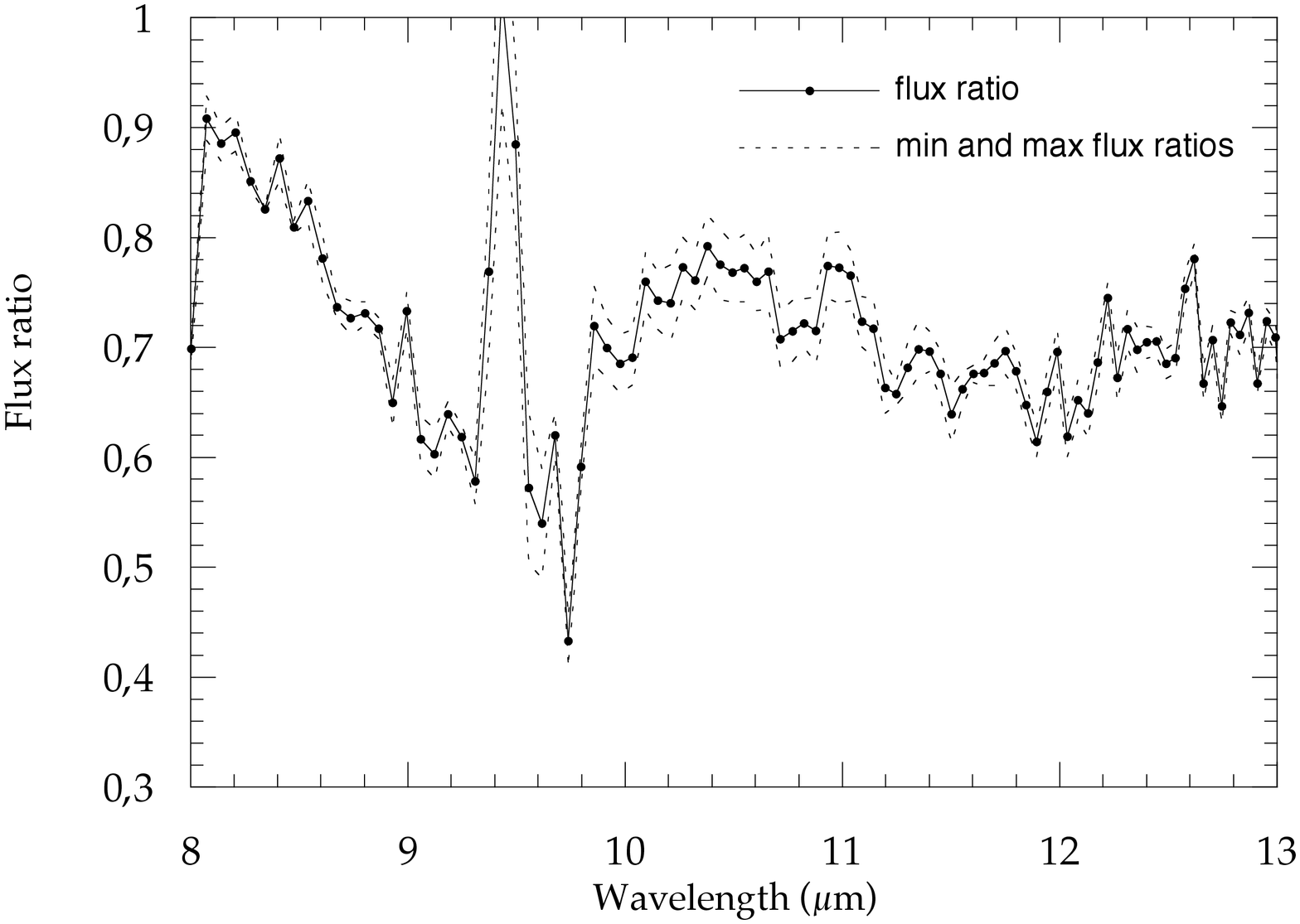}
   \includegraphics[bb=60 58 565 775 , angle=-90, width=8.95cm]{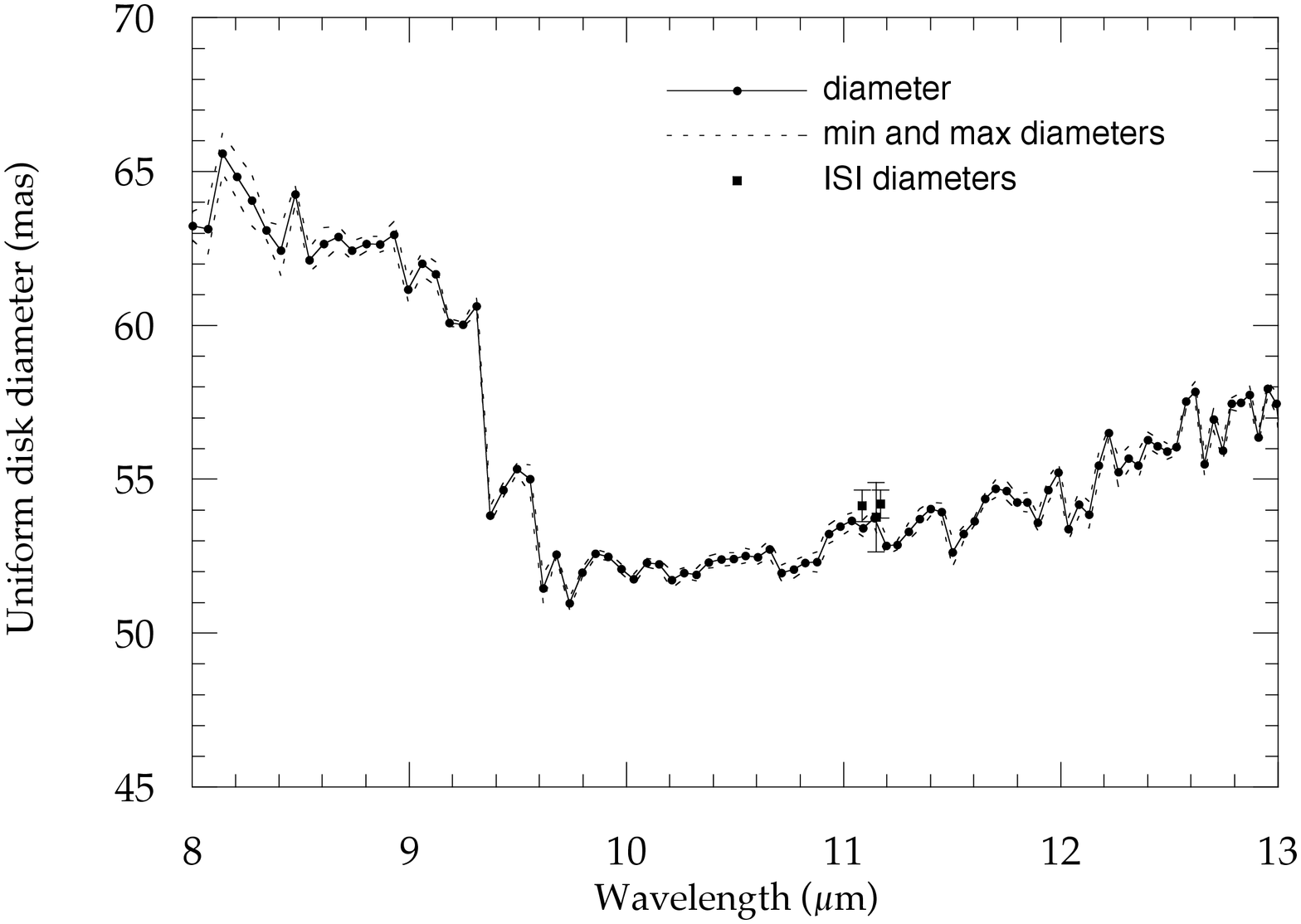}
}   
\vspace{2mm}
     \caption{Fit of the MIDI data by the scaled uniform disk visibility function. Left is the fractional flux of the star $A$. Right is the uniform disk diameter $\O_{UD}$. The MIDI diameters are compared to the ISI diameters at 11.086, 11.149 and 11.171~$\mu$m (Weiner et al. 2003). The 11.149~$\mu$m ISI value is the average of measurements taken at different epochs and the error bar is the standard deviation of these measurements.}
         \label{fig:ratio}
   \end{figure*}

75\% of the fits have $\chi^2$ values smaller than 1 showing that the calibration and the correction of error bars are correct in average. Examples of fits are presented in Fig.~\ref{fig:fits}. Data below 9.18~$\mu$m sample the second lobe of the visibility function of the compact component whereas data above 9.56~$\mu$m sample the first lobe. Confidence in the second lobe arises from the slope of the visibilities which at all wavelengths is positive. Because the source is probably limb-darkened, the fractional flux is likely to be underestimated in the second lobe as a reduced slope due to limb-darkening will be compensated by a lower $A$ with the uniform disk model. Visibilities between 9.18 and 9.56~$\mu$m are difficult to fit as they correspond to the first zero of the visibility function and are very sensitive to errors. They are not used in the following. 

Fig.~\ref{fig:ratio} shows the fractional flux of the compact component and its apparent diameter as a function of wavelength. The graph can be separated into three regions. Below 9.18~$\mu$m, the fractional flux of the compact component is very large. However, because of the bias due to the unknown limb darkening, the fractional flux in this range is probably underestimated and makes its interpretation difficult.
A trough is clearly visible around 9.7~$\mu$m which could be the signature of silicates in the bright resolved dusty environment. A smaller trough is also present between 11 and 12\,$\mu$m which could be due to alumina (Al$_2$O$_3$) -- as it may be present in low-density shells (Sedlmayr \& Kr\"uger 1997) -- or other species. The fractional flux above 10~$\mu$m is roughly constant but significantly smaller than 1 which could possibly be the onset of the signature of the 18\,$\mu$m silicate feature.
The apparent size below 9.56~$\mu$m is comparable to the size measured in the visible probably indicating that the same regions are sampled in these respective wavelength ranges. The optical depth would be large in the mid-infrared and the radiation could be scattered in the visible. Above 9.56~$\mu$m the apparent size steadily increases.

The complex behavior of the apparent size of the star can be most simply understood as a result of wavelength dependent opacity in an extended atmosphere.
\section{Spherical shell model}
\label{model}

\begin{figure*}[]
\hbox{
   \includegraphics[bb=60 58 565 775 , angle=-90, width=8.95cm]{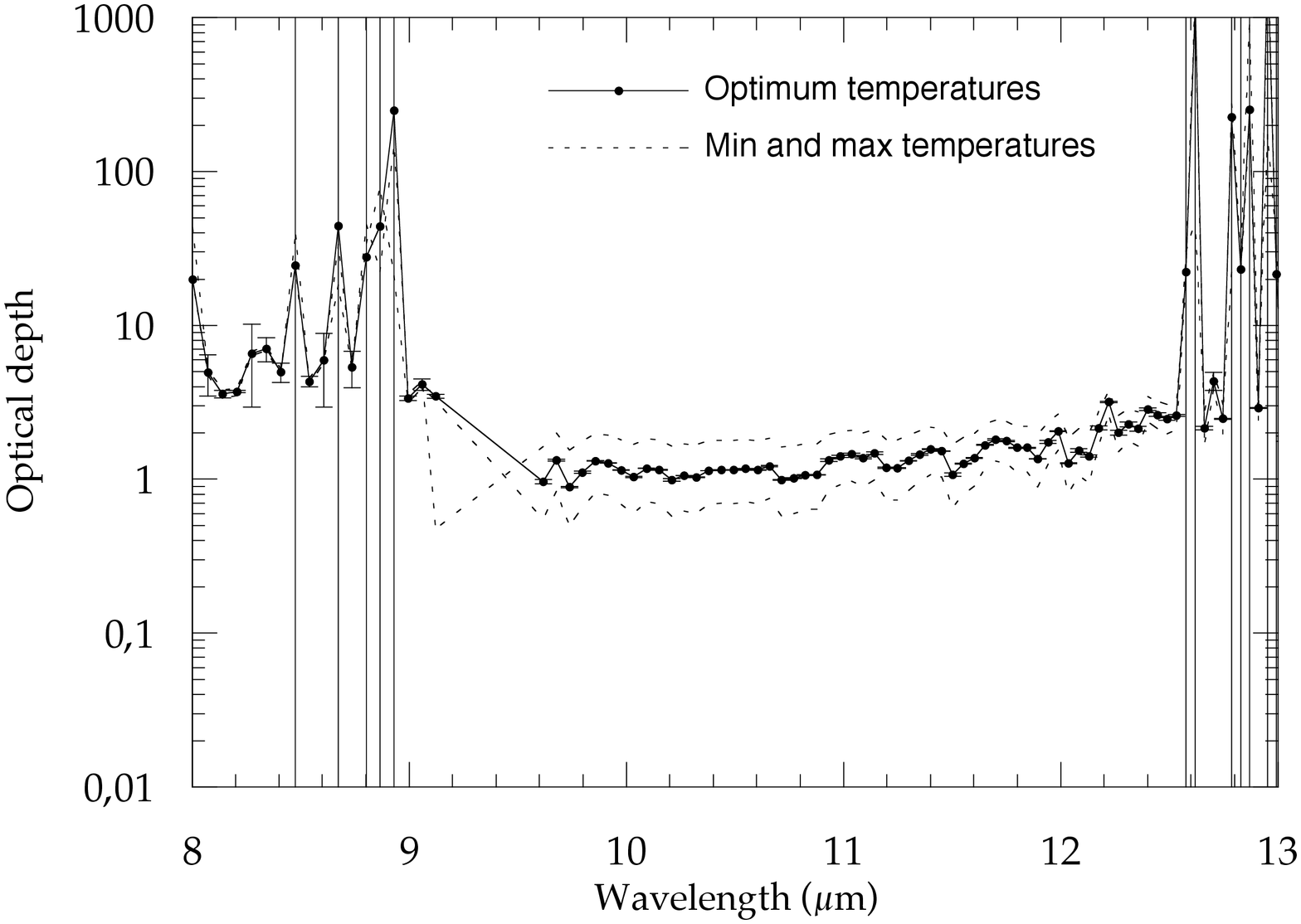}
   \includegraphics[bb=60 58 565 775 , angle=-90, width=8.95cm]{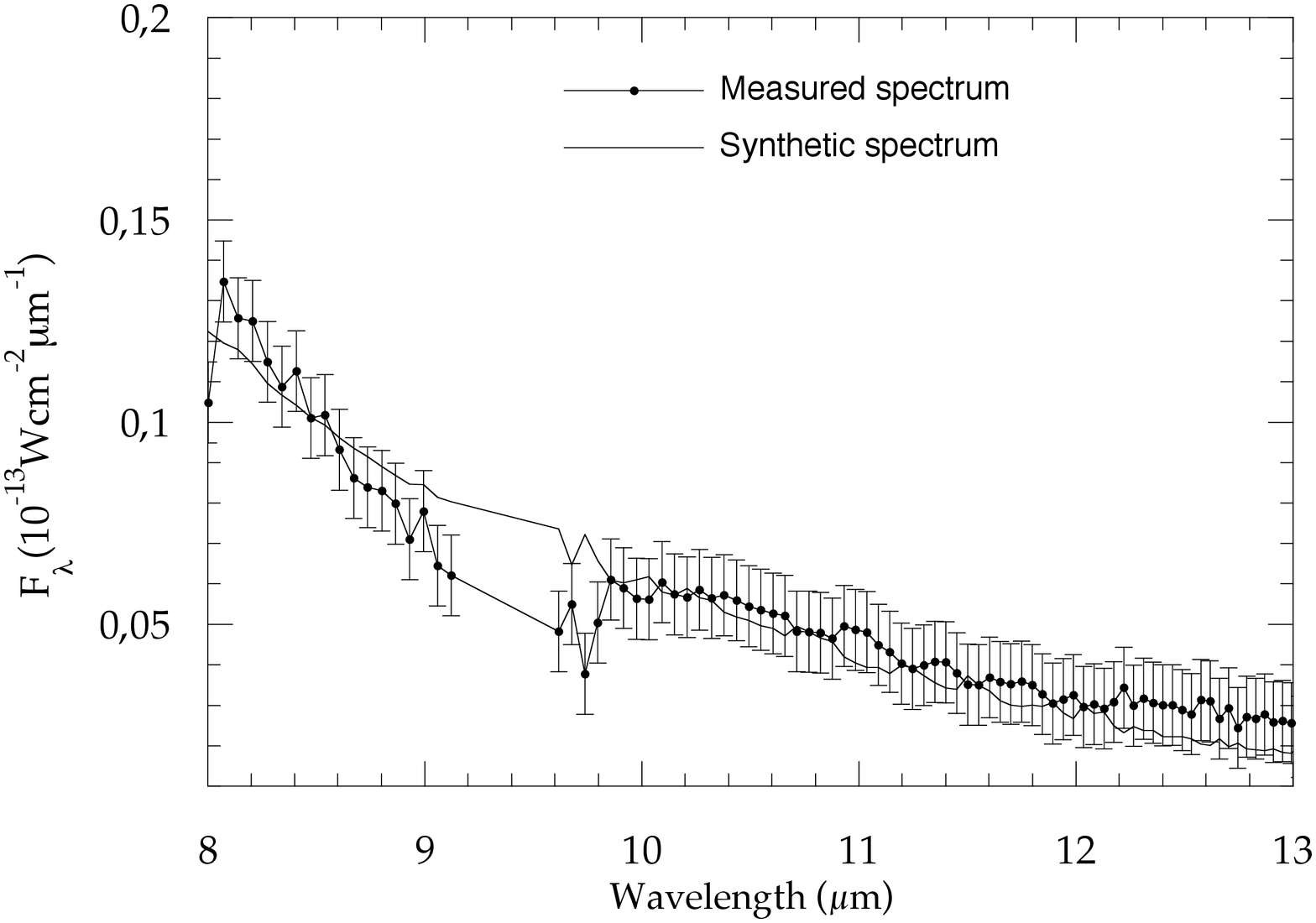}
}   
\vspace{2mm}
     \caption{Result of the fit of the normalized visibilities by the photosphere + shell model. The results are obtained with two independent fits shortward of 9.18~$\mu$m and longward of 9.56~$\mu$m. The left panel is the optical depth of the shells as a function of wavelength. The model spectrum is compared to the measured and normalized spectrum (to account for the dust around the star, see text) in the right panel. The shell temperatures are adjusted to better fit the measured spectrum. }
         \label{fig:tau}
   \end{figure*}

We have shown in the previous section that the apparent diameter of the compact source, the star embedded in the dusty shell, varies with wavelength. In \cite{perrin2004a} we had proposed a model to consistently explain the star diameter difference between the near-infrared (K and L bands) and the mid-infrared (11.15~$\mu$m). The model consisted of a photosphere surrounded by a spherical gaseous shell possibly containing H$_2$O and SiO. This fairly simple model is consistent with the more elaborated model of the MOLsphere of \cite{tsuji2006}. The independent set of data presented in this paper offers an excellent opportunity to test and refine this model. 

The MIDI visibilities have been rescaled  by the reciprocal of the fractional flux factor $A$ of Eq.~\ref{eq:vis} to cancel the visibility loss due to the more extended environment. All wavelengths have then been fitted by the star + spherical shell model whose spatial surface brightness distribution is defined by:

\begin{eqnarray}
I(\lambda,\mu) & = & 
B(\lambda,T_{\star})\mathrm{e}^{-\tau_{\lambda}/\mu} \\ \nonumber 
& & +B(\lambda,T_{\mathrm{shell}})\left[1-\mathrm{e}^{-\tau_{\lambda}/\mu}\right]
\end{eqnarray}
for $\mu \geq \sqrt {1-\left( \O_{\star}/\O_{\mathrm{shell}} \right)^2}$
and:
\begin{equation}
I(\lambda,\mu)=B(\lambda,T_{\mathrm{shell}})\left[1-\mathrm{e}^{-2\tau_{\lambda}/\mu}\right]
\end{equation}
otherwise. \\
$\mu=\cos\theta$ with $\theta$ the angle between the radius vector and the direction to the observer. The parameters of the model are the diameter of the photosphere $\O_{\star}$ and its temperature $T_{\star}$, the diameter and temperature of the shell $\O_{\mathrm{shell}}$ and $T_{\mathrm{shell}}$, and the shell optical depth $ \tau_{\lambda}$, the latter being the only wavelength-dependent parameter of the model. 

We have considered that the photosphere diameter value has been well established from near-infrared observations. This has allowed us to fix two parameters of the model: the diameter of the photosphere and the temperature. We have taken values from \cite{perrin2004a}: $\O_\star=43.71$~mas and $T_\star=3641$~K. The diameter is the average of the two limb-darkened disk diameter estimates of that paper. A consistent limb-darkened diameter has been measured in the H band with visibilities in the first and second lobes (\cite{haubois2006}). In the following, we use this near-infrared limb-darkened diameter as an N-band uniform disk diameter as limb-darkening in this latter wavelength range is supposedly much smaller. This is however a limitation as we do not measure any limb-darkening in the N band and interpret apparent limb-darkening as due to the opacity of the shell. As is discussed later, this is mostly a concern in the lower part of the wavelength range where data are only available in the second lobe of the visibility function. Fixing the photosphere temperature and diameter parameters was anyway necessary as all parameters may be well constrained by the curvature of the visibility function only if the first lobe of the visibility function is well sampled with both high and low visibilities which is not the case with this set of data. Our fixing of these two major parameters at these particular values is however justified as \cite{perrin2004a} have shown that they are in agreement with the bolometric flux and spectral type of the object.
{{
The simultaneous fit of all wavelengths provides the following results:
 }}
\\
 
{ 
{
\begin{tabular}{l}
$\O_{\mathrm{shell}}=62.30\pm0.13$\,mas \\
$T_{\mathrm{shell}}=1598\pm20$\,K \\
$\chi^2=6.29$
\end{tabular} \\
}}

{ {
\noindent In this fit, since the temperature and the size of the photosphere are fixed, the shell temperature reaches a single minimum at optimum value and does not require the total flux of the observed object to be determined in order to mitigate the issue of degeneracy as is the case when the parameters of the photosphere are to be determined. However, a better solution is found if two separate wavelength domains are fitted independantly corresponding to two different altitudes for the shell: shortward of 9.18~$\mu$m and longward of 9.56~$\mu$m as could be guessed from Fig.~\ref{fig:ratio}. But in this case the shell temperatures are not well constrained and it is necessary to force the model spectrum to be consistent with the measured spectrum to get accurate and realistic temperatures. The following optimum parameters have been found for the two domains:}}
\begin{center}
\begin{tabular}{lcl}
$\lambda<9.18~\mu\mathrm{m}$ & & $\lambda>9.56~\mu\mathrm{m}$ \\
$\O_{\mathrm{shell}}=62.50\pm0.50$\,mas & & $\O_{\mathrm{shell}}=57.25\pm0.03$\,mas \\
$T_{\mathrm{shell}}=1570\pm150$\,K & & $T_{\mathrm{shell}}=1520\pm400$\,K \\
\end{tabular} \\
\end{center}
with a global reduced $\chi^2$ of $\chi^2=3.24$. \\

{{
The temperatures found for the two shells are very consistent with the temperature found with the single-shell model. Because the uncertainties are large, it is not possible to discuss the relative values of the shell temperatures. We only consider the two-shell model in the following.
}}

{{
The optical depth is plotted as a function of wavelength in the left panel of Fig.~\ref{fig:tau}. 
Some $\tau_\lambda$ values  are poorly defined in both the lower and upper wavelength sub-ranges. Below $9.18~\mu$m, data have been obtained in the second visibility lobe and this spatial frequency range is very sensitive to limb darkening. In this case the fractional flux may have been underestimated and has led to large visibilities, sometimes higher than the summit of the second lobe. When this happens, the optical depth and the error bar have to be large. }}
Limb darkening in the second visibility lobe also has an impact on the derived diameter. The comparison of the diameter found for the shell below 9.18~$\mu$m and the maximum apparent diameter found in Section~\ref{sec:diameter} may be found surprising as the shell diameter is smaller by 2~mas. This is most probably due to a bias in the fit of the second visibility lobe data by a uniform disk function. The size of the object is very sensitive - to this level of accuracy - to the assumed amount of limb darkening. With the photosphere + shell model, the photosphere is naturally limb darkened when the shell has a non zero optical depth. Visibilities in the first lobe would certainly allow to derive a better estimate of the apparent diameter more in agreement with the derived shell diameter. 
{{
Data taken in the first visibility lobe above 9.56~$\mu$m are far less sensitive to this effect. However a few $\tau_\lambda$ values are large with large error bars. This can be attributed to calibration issues in this upper part of the wavelength range where sensitivity to background is much larger. These values are not to be taken into account for the interpretation of the optical depth.
}}

The apparent diameter found at 11.15~$\mu$m is quite consistent with that measured with ISI by \cite{weiner2000}: respectively $53.75\pm0.32$\,mas versus $54.74\pm0.34$\,mas. The difference may arise from different reasons one of which is that the bandwidth of ISI is by far much narrower than MIDI's, respectively about 2~nm versus 100~nm. The two shells we find are just above the shell of \cite{perrin2004a} whose diameter was 55.78~mas. These altitudes are satisfyingly consistent as the data sets are difficult to compare and could be improved in either case. There is also a good agreement between these shells and the (62.35~mas,1750~K) shell of \cite{verhoelst2006}.

The spectrum of Betelgeuse measured by MIDI is also presented in Fig.~\ref{fig:tau}. More exactely, we have multiplied the spectrum measured by \cite{verhoelst2006} with the same set of MIDI data by the fractional flux of the central object derived in Section~\ref{sec:diameter} and presented in Fig.~\ref{fig:ratio} in order to eliminate the contribution of dust to the spectrum. The error bars have been estimated to $0.01\times10^{-13}$\,Wcm$^{-2}\mu$m$^{-1}$ to account for the uncertainty on the flux ratio. Error bars for the temperatures of the two shells have been determined by matching the model flux uncertainty to that of the measured spectrum. On the same figure, upper and lower limits for the optical depth are plotted (dashed lines). They correpond to the optical depth derived for $\pm 1 \sigma$ values of the temperature as constrained by the spectrum. They give an idea of the probable range of values to be expected for the optical depth as this range is larger than the purely statistical error bar derived from the $\chi^2$ analysis.

\section{Spectroscopic modeling}
\label{spectro}

In this section, we derive the composition of the layer(s) found in
the previous section. As discussed by Scholz~(2001), low spectral resolution visibilities of extended atmospheres with strong molecular opacity cannot be correctly modelled in any other way than by computation of the model and the corresponding visibilities at high resolution (resolving the individual lines), before convolving those resulting visibilities with the instrumental spectral response. Yet, our data set comprises too few baseline coverage and sometimes in a very sensitive spatial frequency range (the second lobe of the visibility  function below $9.56~\mu$m) to perform this direct and rigorous analysis. We therefore first do a comparison of molecular and dust opacities to the empirically derived opacity profile of Sect.~\ref{model} over the N-band, the geometrical model being used as a proxy of the star. This allows the identification of possible constituents and to estimate column densities.  We also perform a direct confrontation between model visibilities and observed visibilities in Sect.~\ref{direct} and study the limits of this approach in the context of our current dataset.\\
In the following, we assume a single layer with a temperature of
1520\,K and reaching from 1.32 to 1.42\,R$_*$. This is a more
realistic representation of the two infinitesimally thin layers found
in the previous section and allows the use of a homogeneous
distribution of the different components.\footnote{A comparison with a
model containing 2 separate layers does not show any significant
differences at the spectral resolution of MIDI and ISO-SWS.}
We have chosen to derive error bars on the column
density based on the observational error in the optical depth determination under
the assumption of a single-temperature layer. Assuming a single
temperature across the N band yielded an uncertainty on this temperature
of only 20\,K in the previous section whose impact on $\tau$ uncertainties is negligible. We did however rescale the error bars on the $\tau$ measurements uniformly in order to obtain a $\chi^2_r = 1$ for the best-fit model.

%
%


\subsection{Possible sources of opacity}

\begin{figure}
\centering
  \resizebox{\hsize}{!}{\includegraphics{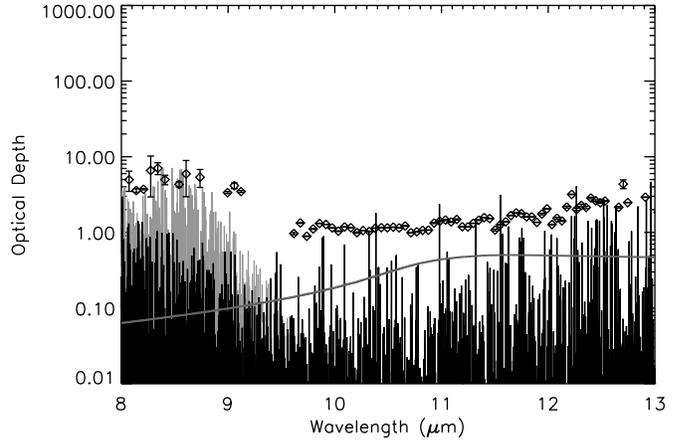}}
  \caption{
  Optical depths observed with MIDI (black diamonds) and high-resolution opacity
  profiles of water (black), SiO (grey) and alumina (solid grey line).
}
  \label{fig:tau_N}
\end{figure}

From the molecules commonly found in CSE's, H$_2$O and SiO are
known to have strong opacity bands in the MIDI wavelength range
(see e.g. Decin~2000). SiO is an absorber below 10~$\mu$m whereas H$_2$O absorbs across the entire range. Other molecules such as silane and ammonia may provide high opacities in carbon-rich stars and supergiants (see Monnier et al. 2000 and references therein) but were not found to be good candidates for our analysis.  In this section, we compare the observed opacity profile
(Fig.~\ref{fig:tau}) with optical depth profiles which we computed for
H$_2$O and SiO at the temperature derived in Sect.~\ref{model}.\\
For H$_2$O, we used the {\sc nasa ames} linelist and partition
function (Partridge \& Schwenke1997), restricted to lines with
$\log{gf - \chi\theta} \ge -9$, with $g$ the ground level statistical
weight, $f$ the oscillator strength, $\theta = 5040/3500$ and $\chi$
the excitation energy in eV. For a justification of the use of this
particular linelist and of the applied cut-off in line strength, we
refer to Cami~(2002), Van~Malderen~(2003) and
Decin~(2000).  For SiO, we used the line list of
Langhoff \& Bauschlicher~(1993), which is the most complete list to
date (Decin 2000). The polynomial expansion of the appropriate
partition function was taken from Sauval \& Tatum~(1984).
Although very high rotational and vibration levels are included in the line list, SiO causes opacity in the N band mostly in its fundamental mode at 8.12~$\mu$m, and
in two hot resonance bands at 8.20 and 8.28~$\mu$m. The microturbulent
velocity was taken to be 3~km\,s$^{-1}$, a typical value for molecules
in AGB stars (Aringer et al. 2002) which is probably also valid for
late-type supergiants.\\
The only dust species which can survive at these high temperatures and
which provides significant opacity at these wavelengths is alumina
(Al$_2$O$_3$).  We used the optical constants measured by
Begemann et al. (1997) and Koike et al. (1995).  A grain size of 0.005\,$\mu$m and a density of 4\,gcm$^{-3}$ were assumed with opacities computed in the Mie approximation (spherical grains).\\
In Fig.~\ref{fig:tau_N}, we show the observed opacities across the N
band, and the high-resolution opacity profiles of H$_2$O, SiO and
alumina. \\
%
%
We see that the high opacities below 9.18\,$\mu$m suggest the presence
of a significant column of SiO. Beyond 9.56\,$\mu$m, the increase of
opacity with wavelength is compatible with the presence of both H$_2$O
and alumina.

\subsection{Modelling the observed opacity profile}
\label{undirect}
\begin{figure}
\centering
  \resizebox{\hsize}{!}{\includegraphics{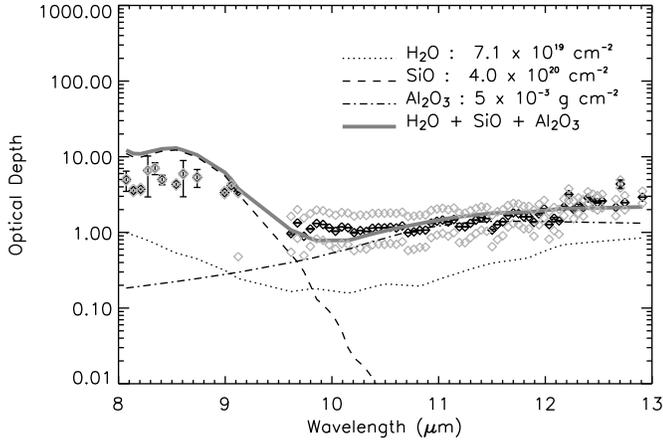}}
  \caption{
  Observed optical depths (black diamonds: optimum temperatures, grey
  diamonds: upper and lower limits on temperatures) compared to the
  best fitting model opacity profile (thick grey solid curve).
}
  \label{fig:SiOH2OAl2O3_tau}
\end{figure}

We attempt a first estimation of the layer composition by
direct comparison of opacities. \\
The computed opacity profiles were rebinned to the resolution of the
MIDI observations (R$\sim$30) and column densities were determined by
searching for the best agreement with the observed optical depths. 
The derived column densities are presented in Table~\ref{tab:cd} and the
result is shown in Fig.~\ref{fig:SiOH2OAl2O3_tau}. To determine realistic
uncertainties on the derived column densities, we rescaled the errors
on the observed optical depths so as to obtain a $\chi^2_r \sim
1$.\\
\begin{table}
\begin{center}
\caption{Derived column densities of H$_2$O, SiO and alumina.}
\label{tab:cd}
\begin{tabular}{l|c}
\hline
\hline
\rule[0mm]{0mm}{4mm} Component &  N [cm$^{-2}$]   \\
\hline
\rule[0mm]{0mm}{4mm} H$_2$O  & $7.1 \pm 4.7 \times 10^{19}$ \\
\rule[0mm]{0mm}{4mm} SiO    & $4.0 \pm 1.1 \times 10^{20}$ \\
\rule[0mm]{0mm}{4mm} Al$_2$O$_3$ $^{\mathrm{a}}\,\,$ & $2.4 \pm 0.5 \times 10^{15}$ \\
\hline
\end{tabular}
\begin{list}{}{}
\item[$^{\mathrm{a}}\,\,$grain size of 0.005\,$\mu$m and density of 4\,gcm$^{-3}$ were assumed]
\item[\,\,\,\,with opacities computed in the Mie approximation (spherical]
\item[\,\,\,\,grains)]
\end{list}
\end{center}
\end{table}

%
%
%
\begin{figure}
\centering
  \resizebox{\hsize}{!}{\includegraphics{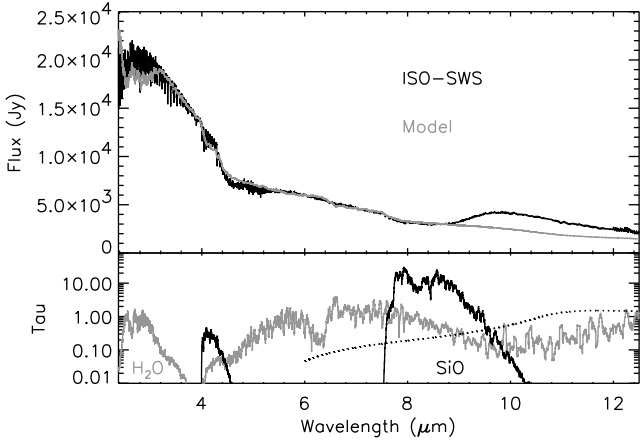}}
  \caption{Upper panel: the near- to mid-IR ISO-SWS spectrum (black) and the
    model with a molecular and dusty (H$_2$O, SiO and alumina) layer (grey). The dusty, detached shell fully resolved by MIDI is not modeled here hence the discrepancy at the silicate emission peak near 9.7\,$\mu$m.
    Bottom panel: opacity profiles of H$_2$O, SiO and alumina (dotted
    line) for the temperatures and column densities in our model.
}
  \label{fig:SiOH2OAl2O3_spec}
\end{figure}

We find that the observed optical depths can be reasonably reproduced
with H$_2$O and SiO column densities of the order of
10$^{20}$\,cm$^{-2}$ and an alumina column density of
5~mg\,cm$^{-2}$. The near- to mid-IR implications of such a layer,
which can be tested against the ISO-SWS spectrum, are shown in Fig.~\ref{fig:SiOH2OAl2O3_spec}.
%
%
%
In general, our model represents the ISO-SWS spectrum reasonably
well. The overestimation of the absorption by H$_2$O at 2.5\,$\mu$m
suggests that our layer temperature may be about 50\,K too low. Still,
this model is in agreement with that proposed by
Verhoelst et al. (2006). 

\subsection{Confrontation with the visibilities}
\label{direct}
Since our approach in the previous section is not rigorously
correct, we test our model against the observed visibilities.

\subsubsection{With the column densities of Sect.~\ref{undirect} and the geometrical model parameters }
\begin{figure}
\centering
  \resizebox{\hsize}{!}{\includegraphics{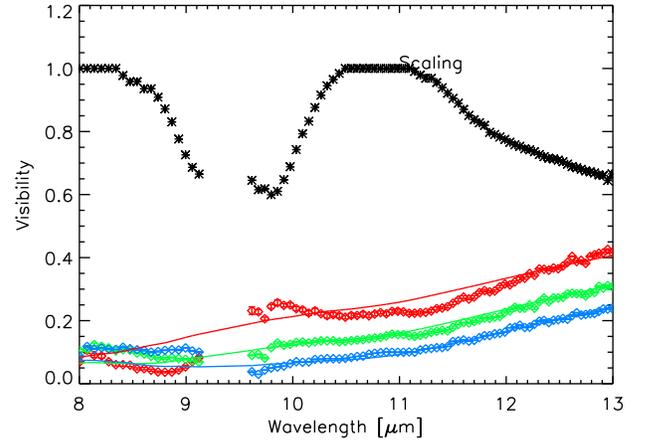}}
  \caption{
Observed and modeled MIDI visibilities are shown in red, green and
blue for the 37.5\,m, 41.5\,m and 44.7\,m projected baselines respectively. The
black crosses show the best-fit scaling factor for each wavelength. 
}
  \label{fig:MIDI_simul}
\end{figure}

We first computed visibilities across the N-band at the MIDI
resolution based on the column densities derived from the $\tau$ fitting in the
previous section and compare these to the observations in
Fig.~\ref{fig:MIDI_simul}. We allowed for a new determination of the
scaling factor due to the flux of the resolved component that takes into account the limb-darkening of the photosphere by the molecular shell. \\
%
%
%
We see that the match in the first lobe of the visibility curve,
i.e. the wavelengths longward of 9.56~$\mu$m, is quite good. The
agreement is much poorer at shorter wavelengths. This is due to
insufficient opacity of the layer at these wavelengths: the SiO lines
(at this layer temperature) are too sparsely spaced so the visibility
is still dominated by photospheric photons, an effect which can't be
overcome by increasing the column density. This is much less a problem
for H$_2$O which has many more closely spaced lines. Moreover the
opacity of our model at longer wavelenghts is dominated by the
continuum opacity of alumina.\\
The scaling function clearly shows the silicate features at 9.7 and
18~$\mu$m in absorption, indicating that the resolved component is, as
expected, the olivine-rich circumstellar dust shell. This is also compatible with absorption by alumina or other species at 11.3$\,\mu$m.

\subsubsection{With column densties and layer diameter as free parameters}
\begin{figure}
\centering
  \resizebox{\hsize}{!}{\includegraphics{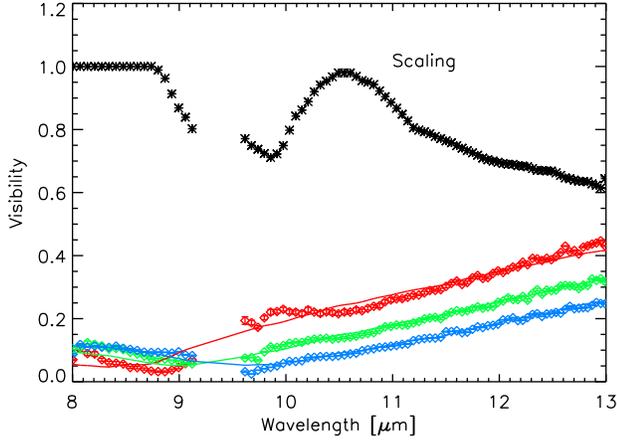}}
  \caption{Best matching model in our grid of 1520\,K layers. Colours and
  symbols as in Fig.~\ref{fig:MIDI_simul}.
}
  \label{fig:bestmodel}
\end{figure}

\begin{figure}
\centering
  \resizebox{\hsize}{!}{\includegraphics{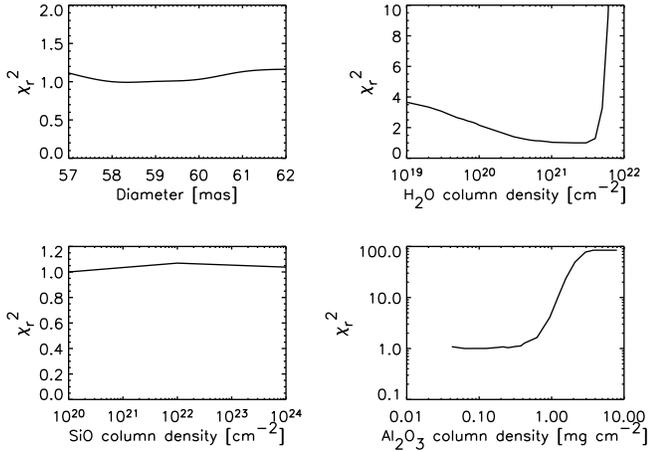}}
  \caption{$\chi^2_r$ (rescaled to 1 for the best model) as a function of the
different parameters. }
  \label{fig:chi2plot}
\end{figure}

\begin{figure}
\centering
  \resizebox{\hsize}{!}{\includegraphics{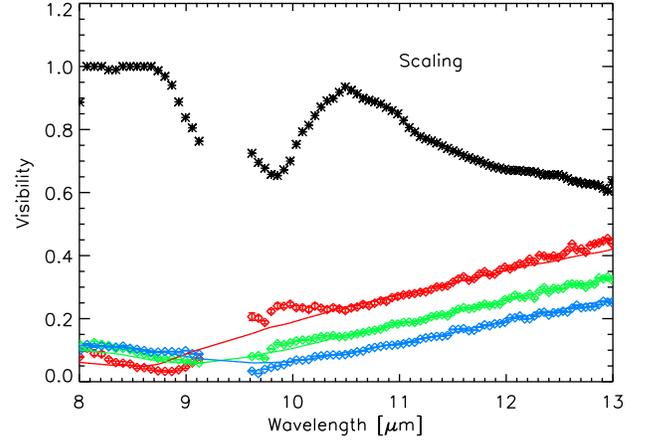}}
  \caption{
Model with a layer at 2000\,K and containing H$_2$O, SiO and
Al$_2$O$_3$. Colours and  symbols as in Fig.~\ref{fig:MIDI_simul}.
}
  \label{fig:fit_MIDI_2000K}
\end{figure}

To improve on our determination of the layer composition, we searched
a grid of models for the best agreement with the observed
visibilities. Our grid ranges from 57 to 62\,mas in layer diameter (1.30 to 1.42\,R$_{\star}$ in layer radius)
with 1\,mas stepsize, $1
\times 10^{19}$ to $1 \times 10^{22}$\,cm$^{-2}$ in H$_2$O column
density with a $1 \times 10^{19,20,21}$ stepsize, $1
\times 10^{20}$ to $1 \times 10^{24}$\,cm$^{-2}$ in SiO column density
and 0.05 to 8\,mg\,cm$^{-2}$ in alumina column density. The lowest
$\chi^2$ is found for a layer of 58\,mas in diameter (1.33\,R$_\star$ in radius), with column densities of    
$3 \times 10^{21}$\,cm$^{-2}$, $1 \times 10^{20}$\,cm$^{-2}$ and
0.1\,mg\,cm$^{-2}$ respectively. This model is shown in
Fig.\,\ref{fig:bestmodel}. 
%
%
%
The $\chi^2_r$ (rescaled to 1 for the best model) as a function of the
different parameters is shown in Fig.~\ref{fig:chi2plot}. We can see
that the water column density is reasonably well constrained. The SiO
has in fact no noticable effect on the low-resolution
visibilities. Alumina can be present, but we find an upper limit of
approximately 1\,mg cm$^{-2}$. In contrast to the empirical modelling
presented in Sect.\,\ref{model}, we find here a very poor
constraint on the layer diameter.  \\
%
%
%
Since a water column density of the order of $10^{21}$\,cm$^{-2}$ is
incompatible with both near-IR -- interferometric and spectroscopic -- observations (too much opacity) and with the
high-resolution ISI observation at 11~$\mu$m (too little opacity), we
may consider a higher layer temperature, e.g. 2000K: this strengthens
the hot resonances of SiO, resulting in lines which are spaced much
closer. The increased SiO opacity below 9.18~$\mu$m could possibly
reduce the required amount of H$_2$O and increase the required column
density of alumina.\\
A fit to the observations of a model with a layer at 2000\,K
containing H$_2$O, SiO and Al$_2$O$_3$ is shown in
Fig.~\ref{fig:fit_MIDI_2000K}. We find the best agreement with column
densities of $7.4 \times 10^{20}$\,cm$^{-2}$, $8.0 \times
10^{22}$\,cm$^{-2}$ and $4.2 \times 10^{-6}$\,g\,cm$^{-2}$ respectively and a layer diameter of 59.13\,mas (1.35\,$R_\star$ in radius).
%
%
%
Indeed, we find that this does increase the SiO column density but
lowers H$_2$O column density not enough and the alumina column too
much. It therefore brings no improvement in the near-IR and
high-resolution 11\,$\mu$m properties of our model.  \\
At this stage we conclude that a rigorous modelling of our data is not possible and that at least a better sampling of the visibility function at smaller spatial frequencies shortward of 9.18\,$\mu$m is required.

\section{Discussion}
\label{discussion}
\subsection{The MOLsphere}
These are the very first interferometric data that allow to study the characteristics of Betelgeuse over a wide mid-infrared spectral range. The only other mid-infrared data available so far have been collected by the ISI interferometer. Comparison of our derived diameters with that of ISI at 11.086, 11.149 and 11.171~$\mu$m (Weiner et al. 2003) is excellent as Figure~\ref{fig:ratio} shows although the ISI spectral resolution is an order of magnitude higher than ours. The strength of the MIDI data is their spectral coverage which allow us to go further in the spatial and spectral understanding of the source.  We derive the optical depth of the environment above the photosphere as a function of wavelength. Absorption is clearly not due to a continuum source of opacity across the N band as the optical depth drops above 9~$\mu$m. Several types of absorbers are required to reproduce the measured optical depth. \\
The ad-hoc model of the MOLsphere is well adapted to capture most of the source characteristics both interferometrically and spectroscopically. Potential constituents are identified. Spectro-interferometric data are well modeled with H$_2$O, Al$_2$O$_3$ and SiO in the MOLsphere. 
 This confirms the presence of dust species which can survive close to the star thanks to their glassy nature - the low opacity of alumina at optical and ultraviolet wavelengths prevents strong heating by the star radiation (Woitke 2006).
Furthermore, the presence of gaseous phase SiO in the MOLsphere strengthens the scenario of \cite{verhoelst2006} for dust formation, whereby alumina may condense first and act as a nucleation site for silicate dust which can condense further from the star where the temperature drops below 1000\,K. The current interferometric data set is not enough to allow a more complex model and we had to assume a geometrically thin shell around the star. In particular as underlined in Section~\ref{sec:diameter} a more complete set of data is needed both at shorter and somewhat longer baselines in order to better measure limb darkening and better measure the optical depth of the shells above the photosphere. This would allow to test this model further and possibly discuss more complex spherical structures and temperature-density radial variations. \\
With this rich data set, it is interesting to review our results in the light of other available studies of the MOLsphere. Our modeling mostly accounts for the ISO-SWS spectrum at a low resolution level. However, even at low resolution the agreement is not perfect (for example between 2.4 and 3.2~$\mu$m) and the too simple description of the MOLsphere in our model is potentially an issue. Apart from this discrepancy, the agreement is globally very good from the upper near-infrared range up to the mid-infrared (the silicate dust infrared excess is not accounted for in our modeling since the dusty environment has been completely resolved out). \\
\cite{ohnaka2004} has modeled a MOLsphere containing water vapor only. Similar parameters to ours were used to model the star ($R_\star=650\,R_\odot$ and $T_{\mathrm{eff}}=3600\,$K) 
but the MOLsphere description is a little different: it extends from the photosphere up to $R_
{\mathrm{mol}}=1.45\,R_\star$. A temperature $T_{\mathrm{mol}}=2050\,$K and a water vapor 
column density of $3\pm1\times10^{20}\,{\mathrm{cm}}^{-2}$ are found. In \cite{tsuji2006}, a slightly different description is used. The MOLsphere is characterized by its inner radius $R_{\mathrm{in}}$ above which the density is decreasing up to the outer radius $R_{\mathrm{out}}$. The same parameters are used to describe the star. Different sets of parameters are investigated for the MOLsphere and the preferred solution is $T_{\mathrm{mol}}=2250\,$K with $R_{\mathrm{in}}=1.31\,R_\star$ and a water vapor column density of $1\times10^{20}$\,cm$^{-2}$. The two models are very consistent as far as water vapor is concerned. Given the different geometries considered by the two authors ($R_{\mathrm{mol}}$ is the upper limit of the MOLsphere in one case whereas $R_{\mathrm{in}}$ is the lower limit in the other one), one expects the Ohnaka radius to be larger than the Tsuji radius which is exactly as they are. The two models define a range of radius for the MOLsphere of 1.31 to 1.45$\,R_\star$ which is exactly the range considered in our spectroscopic model. Column densities are also very consistent to ours.  We however find a lower temperature range although, as previously stated, the uncertainty on this particular parameter is somewhat difficult to define with the current data set. Despite this difference, we conclude the different modelings are quite consistent and we bring evidence for new constituents of the MOLsphere. This is a strong support for the MOLsphere model and we agree with \cite{tsuji2006}Ê that the photospheric origin of all water vapor lines suggested by \cite{ryde2006} is not consistent with both spectroscopic and interferometric data. \\

 It is interesting to compare the results of applying the star-shell model to different stellar types. The earlier near-infrared study of Mira stars (Perrin et al. 2004b) determined model parameters which differ significantly in several respects from the model of $\alpha$~Ori.  The Mira photospheres are cooler by $\sim$400\,K.  The Mira shell radii are twice the stellar radius (compared to $\sim$1.4 for $\alpha$~Ori), and yet the Mira shell model temperatures are similar to or slightly higher than the Betelgeuse shell temperature.  These differences highlight the fact that, although a similar model geometry can be applied to the two cases, they are physically very distinct systems.\\

Obviously, one of the next steps will be to better reproduce the spectra at a level for which there is an agreement between the model and the spectral features at the resolution of MIDI. A better sampling of the visibility function is required for that to reduce the number of a priori inputs to the model (such as the star photospheric radius and temperature) and to determine the intrinsic limb darkening of the photosphere and to allow the more rigorous computation of the model.

\subsection{Mass loss}
There are many questions concerning the mass loss process in stars, and different processes may be at work in different environments.  In a star such as $\alpha$~Orionis, a persistent problem concerns the location of dust nucleation and initial acceleration of mass, since a classical hydrostatic atmosphere fails to offer a suitable location for dust formation. \\
The new evidence reported here, integrated with other recent results, brings together important clues which may guide the way to a preliminary understanding of dust formation in the atmosphere of  $\alpha$~Orionis. The H$_2$O column density we measured can be converted to a total gas column density (assuming $n_\mathrm{H_2O}/n_\mathrm{H} = 10^{-7}$ according to Jennings \& Sada (1998) , and a layer thickness of rougly 10\% of its radius) which yields a mass density of $10^{-13}$\,g\,cm$^{-3}$, which is typical for the wind acceleration/dust condensation regions in sophisticated models (see e.g. Fig. 1 in Woitke  2006). Such a gas distribution, not predicted by static models, is therefore dense enough to support dust formation chemistry.
The low inferred temperature of the molecular shell  ($\sim1600$\,K) is in the range of conditions believed required to permit grain nucleation and growth for one or more species which have particularly favorable thermochemical properties.\\
Finally, a third piece of the puzzle is the tentative identification of emission from a condensed grain material (here identified with alumina) which is associated spectroscopically and spatially with the dense, cool region.\\

 The star-shell model is proving to be a fruitful tool for a preliminary reconnaissance of mass loss systems.  Not surprisingly, the shell parameters found for stars of different types (red supergiants, Mira stars) are quite different.  As it becomes possible to estimate molecular compositions and densities with increasing confidence, the results may be a guide to the likely differing processes at work.\\

Clearly many points in this simple description require study, including the energy balance and equlibrium temperature of the material, the temperature distribution in the shell, and the evolution of chemical equilibrium.  Many questions remain, such as the mechanism for establishing the high density, the dynamics of dust and gas acceleration, and the role of time variations and inhomogeneities.  Spectrally resolved mid-IR interferometery appears well suited to observationally constrain such investigations.




\end{document}